# Results of a campaign to observe outbursts of the dwarf nova CSS 121005:212625+201948

Jeremy Shears, James Boardman, David Boyd, Denis Buczynski, Pavol Dubovský, Juan-Luis González Carballo, Kenneth Menzies, Ian Miller, Roger Pickard, Gary Poyner, Richard Sabo and Richard Sargent

**Abstract**

A monitoring programme of CSS 121005:212625+201948 covering nearly two observing seasons has shown that it is a typical SU UMa dwarf nova, but it has one of the shortest supercycles of its class, at 66.9(6) d. The superoutbursts are interspersed with 3 to 7 short duration (~2 days) normal outbursts each of which are separated by a mean interval of 11 days, but can be as short as 2 days. The most intensively studied superoutburst was that of 2014 November, which lasted 14 days and had an outburst amplitude of >4.8 magnitudes, reaching magnitude 15.7 at its brightest. Time resolved photometry revealed superhumps with a peak-to-peak amplitude of 0.2 magnitudes, later declining to 0.1 magnitude. The superhump period was $P_{sh} = 0.08838(18)$ d. Time resolved photometry was conducted during several other superoutbursts, which gave broadly similar results.

**Introduction**

Dwarf novae are a class of cataclysmic variable star in which a white dwarf primary accretes material from a secondary star via Roche lobe overflow. The secondary is usually a late-type main-sequence star. In the absence of a significant white dwarf magnetic field, material from the secondary is processed through an accretion disc before settling on the surface of the white dwarf. As material builds up in the disc, a thermal instability is triggered that drives the disc into a hotter, brighter state causing an outburst in which the star brightens by several magnitudes. Dwarf novae of the SU UMa family occasionally exhibit superoutbursts which last several times longer than normal outbursts and may be up to a magnitude brighter. During a superoutburst the light curve of an SU UMa star is characterised by superhumps. These are modulations in the light curve which are a few percent longer than the orbital period. They are thought to arise from the interaction of the secondary star orbit with a slowly precessing eccentric accretion disc. The eccentricity of the disc arises because a 3:1 resonance occurs between the secondary star orbit and the motion of matter in the outer accretion disc. For a more detailed review of SU UMa dwarf novae and superhumps, the reader is directed to references (1) and (2).

John Greaves (3) identified CSS 121005:212625+201948 (hereafter CSS 121005) as a likely dwarf nova during his analysis of data from the Catalina Real-time Transient Survey (CRTS) operated as part of the Catalina Sky Survey (CSS) (4). CRTS data show the star varying between 15.5V and fainter than 20.5 V (5) (Figure 1). Greaves noted that its colours reported by GALEX (Galaxy Evolution Explorer) were consistent with it being a CV and that both the Sloan Digital Sky Survey (SDSS; g=16.73,r=16.85) and the POSS/DSS survey had recorded the star in outburst. He pointed out that the CRTS data show that it undergoes very frequent outbursts, being "in outburst as much as not", and thus it might be an ER UMa-type dwarf novae. ER UMa stars are a small subgroup of SU UMa-type dwarf novae, which are characterised by very high outburst frequencies (~4 days) and short (19-48 d) supercycles (the time between consecutive superoutbursts) (6). Greaves noted the similarity in the CRTS light curve between the ER UMa system RZ LMi and CSS 121005. Given the rarity of ER UMa systems (7), Greaves advocated follow-up monitoring of the star to determine its outburst frequency and whether it undergoes superoutbursts.

CSS 121005 lies in Pegasus at 21 26 25.08 +20 19 46.4 (J2000.0). It is also catalogued as GSC2.3 N2NM002562, USNO-B1.0 1103-0582665, GALEX J212625.1+201946 and SDSS J212625.08+201946.3. SDSS Data Release 12 identifies it as a blue object, although no spectrum was recorded (8).

Monitoring of CSS 121005 began in 2012 October following Greaves' announcement. This was intensified during 2014 as part of a campaign supported by the BAA VSS, and publicised via the *baavss-alert* email system (9) and the BAA website (10) with the aim of getting as complete a record of the light curve as possible. This paper presents the photometric results, including detailed analysis of several of the outbursts detected in 2014.

**Observations**

We observed CSS 121005 with small telescopes equipped with CCD cameras as shown in Table 1. Images were dark-subtracted and flat-fielded prior to being measured using differential aperture photometry relative to the AAVSO V-band sequence 14410BXY (11). These data are available from the online databases of the BAA VSS (12) and/or the AAVSO International Database (13). We supplemented our observations with V-band data from the CRTS.

**Analysis of individual outbursts**

As will be discussed later, our intensive campaign revealed a number of short and long outbursts, typical of the behaviour of an SU UMa system. We shall now review the photometry from four long outbursts observed during 2014, the objective of which was to look for superhumps. The observation log is provided in Table 2.

*2014 November outburst*

The best observed long outburst was that of 2014 November, the light curve of which is shown in Figure 2. The system appeared to be brightening from quiescence on the morning (UT) of November 17 (JD 2456978.6) and by the evening it was in full outburst at magnitude 16.5. Careful examination of the light curve suggests that there might have been a slight dip in brightness at this point (JD 2456979.7), which if true might indicate a precursor outburst such as has been seen in other SU UMa systems (see references (14) and (15) for examples in V342 Cam and NN Cam respectively). The following evening (JD 2456980.4) it was brightening again. This led into a plateau phase which lasted about 10 days, with a maximum brightness of 15.7 and a slow decline of ~0.1 magnitude/d. Finally there occurred (JD 2456992) a rapid decline towards quiescence. On JD 2456995 the star was at 18.9, still above quiescence. Thus the total duration of the outburst was ~14 days with an amplitude of at least 4.8 magnitudes (taking quiescence as V >20.5 from the CRTS data). The overall outburst profile is typical of an SU UMa system, with a rapid rise to peak brightness, a plateau with a gentle decline, followed by a rapid drop towards quiescence.

Time series photometry was conducted at various points in the outburst, although none of the runs were particularly long due to the position of the field in the sky at this time of year.

Some of the longer photometry runs are shown in Figure 3. Hump-like features were observed in the light curve on JD 2456982 which continued throughout the plateau phase. We interpret these as superhumps, confirming that CSS 121005 is an SU UMa system. The peak-peak amplitude of the superhumps was 0.2 magnitudes from JD 2456982 to 2458986, declining to 0.1 mag on JD 2456987, and by JD 2456991 superhumps were no longer visible (data not shown). Careful inspection of the photometry in Figure 3 shows very small hump-like features on the descending branch of several of the superhumps. These might be orbital humps or flickering activity commonly seen in dwarf novae.

We performed an ANOVA analysis on the combined photometric data from JD 2456982 to 7, having made a heliocentric conversion and subtracting the mean magnitude, using the *Peranso* software version 2.51 (16). The resulting power spectrum is shown in Figure 4a, which has its highest peak at 11.3150(22) cycles/d, and its 1 cycle/d aliases, which we interpret as the superhump signal corresponding to $P_{sh}$ = 0.08838(18) d. The error estimates are derived using the Schwarzenberg-Czerny method (17). Several other statistical algorithms in Peranso gave the same value of $P_{sh}$, including Scargle's method (18) as optimised by Horne and Baliunas (19).

Removing the superhump signal from the power spectrum in Figure 4(a) leaves only very weak signals. We could find no evidence for a signal relating to the small orbital hump-like features in the descending branch of the superhumps. A phase diagram of the combined data folded on $P_{sh}$ is shown in Figure 5, where two cycles are shown for clarity.

To study the superhump behaviour further, we extracted the times of each sufficiently well-defined superhump maximum from the individual light curves according to the Kwee and van Woerden method (20) using *Peranso*. Times of 12 superhump maxima were found and are shown in Table 3 along with the error estimates from the Kwee and van Woerden method. Following a preliminary assignment of superhump cycle numbers to these maxima, we obtained the following unweighted least-squares linear superhump maximum ephemeris for the interval JD 2456982 to 7 (covering 57 superhump cycles):

$$HJD_{Max} = 2456982.5509(23) + 0.08830(1) \times E \qquad \text{Equation 1}$$

The superhump period $P_{sh}$ = 0.08830(1) is consistent with the value derived from the ANOVA analysis. The observed minus calculated (O–C) residuals relative to this ephemeris are plotted in the bottom panel of Figure 2. A quadratic fit to the data suggests that the superhump period changed over the interval investigated with $dP_{sh}/dt = -2.67(27) \times 10^{-4}$, which is consistent with the evolution of $P_{sh}$ in SU UMa systems reported by *Kato et al.* (21)

*2014 September outburst*

The outburst was detected on 2014 September 6 (JD 2456907.4) and its light curve shown in Figure 6(a) is similar to the November outburst, although less well observed. It lasted at least 12 days and was magnitude 15.7 at its brightest. Once again none of the photometry runs was particularly long, but superhumps were present during the plateau phase (data not shown). Whilst we were able to measure maximum times for a few superhumps, we considered these to be too few for a meaningful analysis, so we carried out period analysis instead. Thus, we combined the data from this phase, between JD 2456910.4 and 2456914.4, and performed an ANOVA analysis in the same way as previously. The resulting power spectrum (Figure 4(b)) has its highest peak at 11.3320(26) cycles/day, and 1 cycle/d aliases, corresponding to $P_{sh} = 0.08825(40)$ d, which is consistent with the value measured in 2014 November.

*2014 July outburst*

This outburst was detected on 2014 July 1 and again lasted at least 12 days. At its brightest it was magnitude 15.9, but it might have been brighter during the first part of the plateau phase which was missed (Figure 6(b)). Again photometry runs were short, mainly due to the short nights of summer, but hump-like features were visible. As rather few superhump times could be measured, we performed an ANOVA analysis of the combined data from JD 2456845.4 to 2456850.6. This gave the power spectrum in Figure 5(c), with its highest peak at 11.32(8) cycles/day, and 1 cycle/d aliases, corresponding to $P_{sh} = 0.0883(6)$ d, again consistent with the $P_{sh}$ values measured in 2014 November and September.

*2014 April outburst*

This outburst occurred near the beginning of the observing season, before the BAA VSS campaign was launched, and was poorly observed, with no significant time-series photometry runs. The light curve in Figure 6(c) shows that the outburst lasted at least 11 days and the star was magnitude 15.8 at its brightest. The similarity of the profile of the outburst to the other three outbursts leads us to conclude that this too was a superoutburst.

**Outburst frequency**

The long term light curve between 2013 July 6 to 2015 Feb 3 is shown in Figure 7 based on data from the authors and the CRTS. The effect of the increased sampling rate from around JD2456840, as a result of the BAA VSS campaign gaining momentum, can clearly be seen. A multiplicity of outbursts is present and the details of 32 are presented in Table 4, where an outburst is identified as a brightening to magnitude 17.0 or brighter. A total of 7 long outbursts (each >8 days, but typically 12 to 14 days) were detected, including the 4 superoutbursts discussed above. We therefore interpret all these as superoutbursts. The final event, in 2015 January/February, was only partially observed before observations were impossible as the field slipped from view at the end of the season. However, its duration indicates that it was a superoutburst.

We assigned preliminary cycle numbers to each superoutburst and performed an unweighted least squares linear fit to the times of superoutburst. From this, the linear ephemeris for superoutburst is:

$$JD_{max} = 2456509(3) + 66.9(6) \times E \qquad\qquad \text{Equation 2}$$

This suggests a superoutburst interval, or supercycle length, of 66.9(6) days. We also measured the times between 5 successive superoutbursts and found an average of 69.4(21) days, where the error is the standard deviation.

The corresponding O-C diagram from the ephemeris in equation 2 is shown in Figure 8 and it is clear that while there is some coherence in the outburst timings, there is also some scatter in the individual outburst intervals. The O-C residuals range over +/- 5 days, or 7% of the supercycle, which is remarkably consistent for a quasi-periodic phenomenon.

The remaining 25 outbursts in Table 4 are very short, typically ~2 days, which we interpret as normal outbursts. Time series photometry was conducted during the 2013 Nov (JD 2456965.4) short outburst which showed no superhumps, again confirming that this was a normal outburst (data not shown). Given the short duration of the normal outbursts it is likely that several have been missed, especially before the BAA VSS campaign gained traction. The mean interval between successive normal outbursts was 11.5 days with a standard deviation of 5.4 days and the shortest interval was 2 days. The number of normal outbursts between successive superoutbursts ranged between 3 and 7.

**Discussion**

The supercycle of CSS 121005 of ~67 days places it at the lower end of the range for SU UMa systems, where one of the shortest currently known is BF Ara with a supercycle of 84 days ($P_{sh} = 0.089$ d) (22). However, it is not quite as short as the known ER UMa systems which, as we saw earlier, is 19-58 days. The search for "transition" objects, if such things exist, between typical SU UMa systems and ER UMa stars has so far drawn a blank and unfortunately CSS 121005 is not a contender. The superhump periods, and therefore the orbital periods, of ER UMa systems are generally <0.07 d, so are among the shortest seen for SU UMa stars and overlap those of the WZ Sge sub-class. For CSS 121005 the $P_{sh} = 0.088$ d, which is longer than the majority of SU UMa systems, but is still typical of the class.

A further inconsistency between the observed behaviour of CSS 121005 and the known ER UMa systems is that the latter spend a large amount of time, 30 to 45%, in superoutburst. By contrast, CSS 121005 spends only 20% of the time in this active state, although this value is still greater than the majority of normal SU UMa systems. Also the mean decline rate in the plateau phase (0.1 mag/d) is more typical of SU UMa systems (23) rather than the extremely small value of ~0.04 mag/d in ER UMa systems (24).

Recently, as a result of many new dwarf nova being revealed in the OGLE-III Galactic Disc Fields, Mróz et al. (25) have noted that the number of detected short supercycle period dwarf novae is becoming large enough to question the validity of the ER UMa classification as being separate from SU UMa type variables. However, Kato et al. (26) have suggested some problems with the analysis of Mróz et al. (25) and are probably of the SS Cyg subtype. The results from the increasing number of synoptic photometric studies will surely shed further light on this.

The results of this campaign once again demonstrate the value of intensive and co-ordinated monitoring of cataclysmic variables by amateur astronomers possessing relatively simple equipment, complemented with time-resolved photometry at multiple longitudes during outbursts.

**Conclusions**

A monitoring programme of CSS 121005 covering two observing seasons has shown that it has a relatively well-defined superoutburst period of 66.9(6) d. The superoutburst are interspersed with 3 to 7 short duration (~2 days) normal outbursts each of which are separated by a mean interval of 11 days, but can be as short as 2 days.

The most intensively studied superoutburst was in 2014 November, which lasted 14 days and had an outburst amplitude of >4.8 magnitudes, reaching magnitude 15.7 at its brightest.

Time resolved photometry revealed superhumps with a peak-to-peak amplitude of 0.2 magnitude, later declining to 0.1 magnitude. The superhump period was $P_{sh} = 0.08838(18)$ d.

Photometry was conducted during several other superoutbursts, which gave broadly similar results. Superhumps were observed in the 2014 July and September superoutbursts with $P_{sh} = 0.08834(59)$ d and $0.08825(40)$ d respectively, which values are consistent with the 2014 November event.

CSS 121005 has one of the shortest superoutburst periods of typical SU UMa systems, but our results confirm it is not a member of the ER UMa family of dwarf novae.

## Acknowledgements

The authors owe a debt of gratitude to John Greaves, who first brought this interesting star to our attention and generously shared the fruits of his personal research, which triggered this project. This research also made use of data from the Catalina Sky Survey (the CSS survey is funded by the National Aeronautics and Space Administration under Grant No. NNG05GF22G issued through the Science Mission Directorate Near-Earth Objects Observations Program; the CRTS survey is supported by the U.S. National Science Foundation under grants AST-0909182 and AST-1313422) and the Sloan Digital Sky Survey (funding for the SDSS has been provided by the Alfred P. Sloan Foundation, the National Science Foundation, the U.S. Department of Energy, the National Aeronautics and Space Administration, the Japanese Monbukagakusho, the Max Planck Society, and the Higher Education Funding Council for England). We used SIMBAD and Vizier, operated through the Centre de Données Astronomiques (Strasbourg, France) and the NASA/Smithsonian Astrophysics Data System.

Shears and Poyner thank the Department of Cybernetics at the University of Bradford, UK, for the use of the Bradford Robotic Telescope (BRT), located at the Teide Observatory on Tenerife in the Canary Islands. Shears also thanks the AAVSO for the use of the SRO-50 telescope which is part of the AAVSOnet facility.

The authors thank the two referees for their helpful and constructive comments that have improved the paper.

## Addresses

Shears: "Pemberton", School Lane, Bunbury, Tarporley, Cheshire, CW6 9NR, UK [bunburyobservatory@hotmail.com]

Boardman: De Soto, WI, USA [j_boardmanjr@yahoo.com]

Boyd:  West Challow, Wantage, Oxon, OX12 9TX, UK [davidboyd@orion.me.uk]

Buczynski: Portmahomac, near Tain, Ross-shire, IV20 1RD, UK [buczynski8166@btinternet.com]

Dubovský: Vihorlat Observatory Humenne, Slovakia [var@kozmos.sk]

González Carballo: Observatorio Cerro del Viento MPC I84, Badajoz, Spain [struve1@gmail.com]

Menzies: Framingham, MA 01701, USA [kenmenstar@gmail.com]

Miller: Ilston, Swansea, SA2 7LE, UK [furzehillobservatory@hotmail.com]

Pickard: Shobdon, Leominster, Herefordshire. HR6 9NG, UK [roger.pickard@sky.com]

Poyner: Kingstanding, Birmingham, B44 0QE, UK [garypoyner@blueyonder.co.uk]

Sabo: Bozeman, MT 59718, USA [rsabo333@gmail.com]

Sargent: Upton, Chester, CH2 2JB, UK [uptonobservatory@hotmail.co.uk]

**References and notes**

1. *Hellier C., Cataclysmic Variable Stars: How and why they vary, Springer-Verlag (2001).*

2. *Warner B., Cataclysmic Variables, Cambridge University Press (2003).*

3. *Greaves, J. Personal communication via email, 2012 October 9.*

4. *Drake, A.J. et al., ApJ, 696, 870 (2009).*

5. *The CSS light curve can be seen at: http://nesssi.cacr.caltech.edu/catalina/20121005/1210051211094121504p.html.*

6. *Kato T. et al., Frontiers Science Series No. 26., p.45 (1999). Available online at http://arxiv.org/abs/1301.3202.*

7. *Otulakowska-Hypka M. et al., MNRAS, 429, 868-880 (2013) list 5 ER UMa systems: ER UMa, V1159 Ori, RZ LMi, DI UMa, IX Dra.*

8. *SDSS DR 12 data at: http://skyserver.sdss.org/public/en/tools/quicklook/quicksummary.aspx?id=0x112d1821c1030199&spec=.*

9. *http://www.britastro.org/vss/alert.htm.*

10. *http://www.britastro.org/.*

11. *The sequence is available at http://www.aavso.org/vsp/.*

12. *http://britastro.org/vssdb/.*

13. *http://www.aavso.org/aavso-international-database.*

14. *Shears J.H. et al., New Astronomy, 16, 311-316 (2011).*

15. *Shears J. et al., JBAA, 121, 355-362 (2011).*

16. *Vanmunster T., http://www.peranso.com/.*

17. *Schwarzenberg-Czerny A., Mon. Not. Royal Astron. Soc., 253, 198 (1991).*

18. *Scargle J.G., ApT, 263, 835 (1982).*

19. *Horne J.H. and Baliunas S.L., ApJ, 302, 757 (1986).*

20. *Kwee K. and van Woerden H., Bulletin of the Astronomical Institutes of the Netherlands, 12, 464 (1956).*

21. *Kato T. et al., PASJ, 61, S395-S616 (2009).*

22. *Kato T., et al., MNRAS, 341, 901-908 (2003).*

23. *Kato T. et al., A&A, 395, 542 (2002).*

24. *Kato T. and Kunjaya C., PASJ, 47, 163 (1995).*

25. *Mróz P. et al, Acta Astronomica, 63, 135-158 (2013).*

26. *Kato T. et al, PASJ, 66, 30 (2014).*

| Observer | Telescope | CCD |
|---|---|---|
| Boardman | 0.35m SCT | SBIG STT-1603-ME |
| Boyd | 0.35m SCT | Starlight Xpress SXVR-H9 |
| Buczynski | 0.35m SCT | SBIG ST9XE |
| Dubovský | 0.35 m SCT | Moravian Instruments G2-1600 |
| Menzies | 0.317 m modified Harmer-Wynne Cassegrain | SBIG STL-6303E |
| González Carballo | 0.2 m SCT | Atik 16HR |
| Miller | 0.35 m SCT | Starlight Xpress SXVR-H16 |
| Pickard | 0.35 m SCT | Starlight Xpress SXVF-H9 |
| Poyner | 0.35 m SCT (Bradford Robotic Telescope) | FLI MicroLine fitted with a E2V CCD47-10 |
| Sabo | 0.43 m reflector | SBIG STL-1001 |
| Sargent | 0.2 m SCT | Starlight Xpress SXVF-H9 |
| Shears | 0.28 m SCT (Bunbury) | Starlight Xpress SXVF-H9 |
|  | 0.35 m SCT (Bradford Robotic Telescope) | FLI MicroLine fitted with a E2V CCD47-10 |
|  | 0.5 m reflector (Sonoita Research Observatory, SRO-50) | SBIG STL 6303 with V-filter |

Table 1: Instrumentation used

CCDs were unfiltered with the exception of the Sonoita Research Observatory

| Date (UT) | Start time (JD) | End time (JD) | Duration (h) | Observer |
|---|---|---|---|---|
| April 30 | 2456777.900 | 2456777.936 | 0.9 | Sabo |
| May 1 | 2456778.892 | 2456778.939 | 1.1 | Sabo |
| July 7 | 2456845.506 | 2456845.597 | 2.2 | Miller |
| July 8 | 2456847.468 | 2456847.581 | 2.7 | Miller |
| July 9 | 2456848.461 | 2456848.602 | 3.4 | Pickard |
| July 9 | 2456848.485 | 2456848.596 | 2.7 | Miller |
| July 11 | 2456849.753 | 2456849.922 | 4.1 | Sabo |
| July 11 | 2456850.445 | 2456850.608 | 3.9 | Pickard |
| July 13 | 2456851.754 | 2456851.920 | 4.0 | Sabo |
| July 13 | 2456852.430 | 2456852.574 | 3.5 | Pickard |
| August 23 | 2456892.509 | 2456892.593 | 2.0 | Miller |
| August 23 | 2456892.559 | 2456892.622 | 1.5 | Pickard |
| August 23 | 2456893.376 | 2456893.424 | 1.2 | Sargent |
| August 23 | 2456893.426 | 2456893.485 | 1.4 | Pickard |
| September 6 | 2456907.388 | 2456907.459 | 1.7 | Miller |
| September 9 | 2456910.348 | 2456910.458 | 2.6 | Sargent |
| September 10 | 2456911.357 | 2456911.448 | 2.2 | Sargent |
| September 10 | 2456911.377 | 2456911.574 | 4.7 | Miller |
| September 11 | 2456912.372 | 2456912.445 | 1.6 | Miller |
| September 11 | 2456912.390 | 2456912.449 | 1.4 | Sargent |
| September 12 | 2456913.348 | 2456913.439 | 2.2 | Miller |
| September 13 | 2456914.366 | 2456914.418 | 1.2 | Miller |
| October 13 | 2456943.679 | 2456943.833 | 3.7 | Sabo |
| October 15 | 2456945.633 | 2456945.808 | 4.2 | Sabo |
| October 17 | 2456947.637 | 2456947.912 | 6.6 | Sabo |
| November 3 | 2456965.376 | 2456965.414 | 0.9 | Boyd |
| November 3 | 2456965.412 | 2456965.438 | 0.6 | Pickard |
| November 3 | 2456965.474 | 2456965.542 | 1.6 | Menzies |
| November 4 | 2456965.633 | 2456965.789 | 3.7 | Sabo |
| November 4 | 2456966.175 | 2456966.384 | 5.0 | Dubovský |
| November 4 | 2456966.308 | 2456966.406 | 2.4 | Pickard |
| November 5 | 2456966.517 | 2456966.668 | 3.6 | Boardman |
| November 5 | 2456967.166 | 2456967.226 | 1.6 | Dubovský |
| November 5 | 2456967.255 | 2456967.437 | 4.4 | Pickard |
| November 18 | 2456979.720 | 2456979.773 | 1.3 | Sabo |
| November 18 | 2456980.256 | 2456980.414 | 3.8 | Miller |
| November 20 | 2456982.274 | 2456982.374 | 2.4 | Shears |
| November 20 | 2456982.441 | 2456982.610 | 4.1 | Menzies |
| November 21 | 2456982.517 | 2456982.626 | 2.6 | Boardman |
| November 21 | 2456983.425 | 2456983.588 | 3.9 | Menzies |
| November 22 | 2456984.245 | 2456984.282 | 0.9 | Pickard |
| November 22 | 2456984.304 | 2456984.332 | 0.7 | Shears |
| November 23 | 2456985.222 | 2456985.391 | 4.1 | Shears |
| November 23 | 2456985.231 | 2456985.396 | 4.0 | Pickard |
| November 23 | 2456985.251 | 2456985.426 | 4.2 | Miller |
| November 24 | 2456986.164 | 2456986.267 | 2.5 | Dubovský |
| November 24 | 2456986.225 | 2456986.410 | 4.4 | Pickard |
| November 24 | 2456986.234 | 2456986.372 | 3.3 | Miller |
| November 24 | 2456986.237 | 2456986.400 | 3.9 | Boyd |
| November 25 | 2456987.244 | 2456987.437 | 4.6 | Buczynski |
| November 26 | 2456987.501 | 2456987.603 | 2.4 | Boardman |
| November 26 | 2456988.238 | 2456988.286 | 1.2 | Buczynski |
| November 28 | 2456989.503 | 2456989.538 | 0.8 | Boardman |
| November 29 | 2456991.153 | 2456991.337 | 4.4 | Dubovský |
| November 30 | 2456992.274 | 2456992.384 | 2.6 | Pickard |

Table 2: Log of time series photometry in 2014

| Superhump cycle number | Time of maximum (HJD) | Measurement error (d) | O-C (d) |
|---|---|---|---|
| 0 | 2456982.5454 | 0.0069 | -0.0055 |
| 10 | 2456983.4326 | 0.0054 | -0.0013 |
| 11 | 2456983.5231 | 0.0049 | 0.0009 |
| 31 | 2456985.2897 | 0.0045 | 0.0015 |
| 31 | 2456985.2922 | 0.0059 | 0.0040 |
| 32 | 2456985.3796 | 0.0049 | 0.0031 |
| 32 | 2456985.3802 | 0.0069 | 0.0037 |
| 42 | 2456986.2592 | 0.0053 | -0.0003 |
| 42 | 2456986.2593 | 0.0068 | -0.0002 |
| 42 | 2456986.2594 | 0.0061 | -0.0001 |
| 43 | 2456986.3481 | 0.0077 | 0.0003 |
| 57 | 2456987.5768 | 0.0095 | -0.0072 |

Table 3: Superhump maximum times during the 2014 November superoutburst

| Detection date (UT) | JD | Outburst type | Superoutburst cycle number | Magnitude at maximum | Duration (d) | Time since previous superoutburst (d) | Time since previous normal outburst (d) |
|---|---|---|---|---|---|---|---|
| 2013 August 5 | 2456509.6 | Super | 0 | 16.1 | >12 | | |
| 2013 September 14 | 2456550.4 | Normal | | 16.8 | | | |
| 2013 September 28 | 2456563.5 | Normal | | 17.3 | | | 13.1 |
| 2013 October 5 | 2456571.4 | Normal | | 16.7 | | | 7.9 |
| 2013 October 14 | 2456580.4 | Super | 1 | 15.7 | >11 | 70.8 | |
| 2013 November 10 | 2456607.3 | Normal | | 16.7 | | | |
| 2013 November 22 | 2456619.3 | Normal | | 16.2 | | | 12.0 |
| 2013 December 3 | 2456630.4 | Normal | | 16.1 | | | 11.1 |
| 2013 December 19 | 2456646.3 | Normal | | 16.5 | | | 15.9 |
| 2013 December 31 | 2456658.2 | Normal | | 17.0 | | | 11.9 |
| ← | | | *Seasonal gap* | | | | → |
| 2014 April 25 | 2456772.9 | Super | 4 | 15.8 | >11 | | |
| 2014 May 14 | 2456792.3 | Normal | | 16.6 | | | |
| 2014 May 28 | 2456805.9 | Normal | | 17.1 | | | 13.6 |
| 2014 June 23 | 2456831.5 | Normal | | 16.2 | | | 25.6 |
| 2014 July 1 | 2456839.9 | Super | 5 | 15.9 | >13 | 67.0 | |
| 2014 July 21 | 2456859.5 | Normal | | 16.8 | ~2 | | |
| 2014 August 3 | 2456872.5 | Normal | | 16.4 | ~2 | | 13.0 |
| 2014 August 22 | 2456892.4 | Normal | | 15.9 | ~2 | | 19.9 |
| 2014 September 6 | 2456907.4 | Super | 6 | 15.7 | >12 | 67.5 | |
| 2014 September 29 | 2456930.4 | Normal | | 17.9 | | | |
| 2014 October 8 | 2456939.4 | Normal | | 16.3 | | | 9.0 |
| 2014 October 15 | 2456945.7 | Normal | | 17.1 | | | 6.3 |
| 2014 October 17 | 2456947.6 | Normal | | 15.9 | ~2 | | 1.9 |
| 2014 October 21 | 2456951.8 | Normal | | 17.4 | ~2 | | 4.2 |
| 2014 November 3 | 2456965.4 | Normal | | 15.7 | ~2 | | 13.6 |
| 2014 November 10 | 2456971.9 | Normal | | 16.6 | 2 to 3 | | 6.5 |
| 2014 November 17 | 2456979.4 | Super | 7 | 15.7 | ~14 | 72.0 | |
| 2014 December 7 | 2456999.3 | Normal | | 17.5 | | | |
| 2014 December 19 | 2457011.3 | Normal | | 17.7 | | | 12.0 |
| 2014 December 28 | 2457020.2 | Normal | | 17.2 | | | 8.9 |
| 2015 January 10 | 2457033.2 | Normal | | 16.6 | | | 13.0 |
| 2015 January 26 | 2457049.3 | Super | 8 | 15.9 | >8 | 69.9 | |

Table 4: Outbursts observed between 2013 August and 2015 February

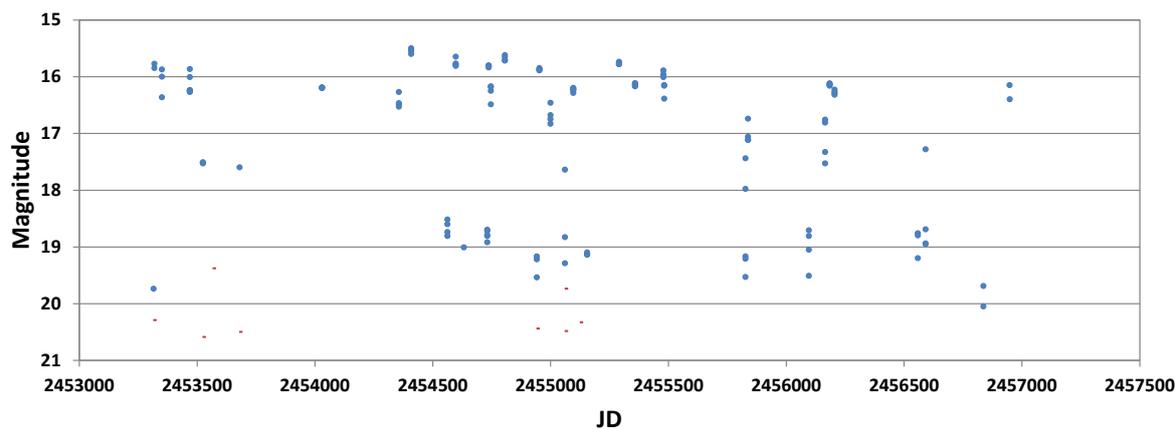

Figure 1: CRTS data on CSS 121005 from 2004 Nov to 2014 Oct

Blue data are V measurements; red dashes are "fainter than" observations

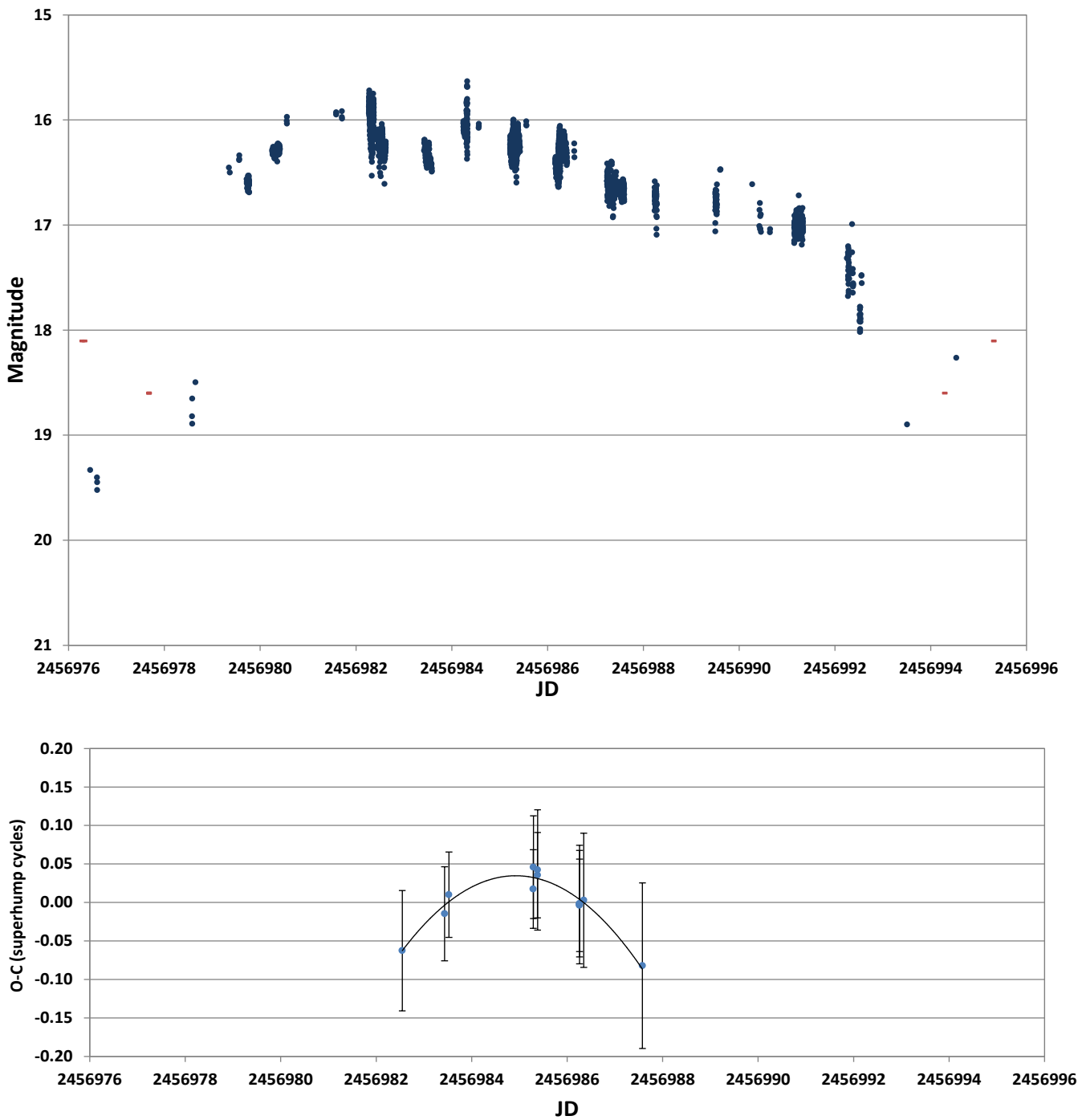

Figure 2: The 2014 November superoutburst

**Top:** outburst light curve. Blue data are unfiltered CCD measurements; red dashes are "fainter than" observations. Bottom: O-C diagram of the superhump times of maximum

*Accepted for publication in the Journal of the British Astronomical Association*

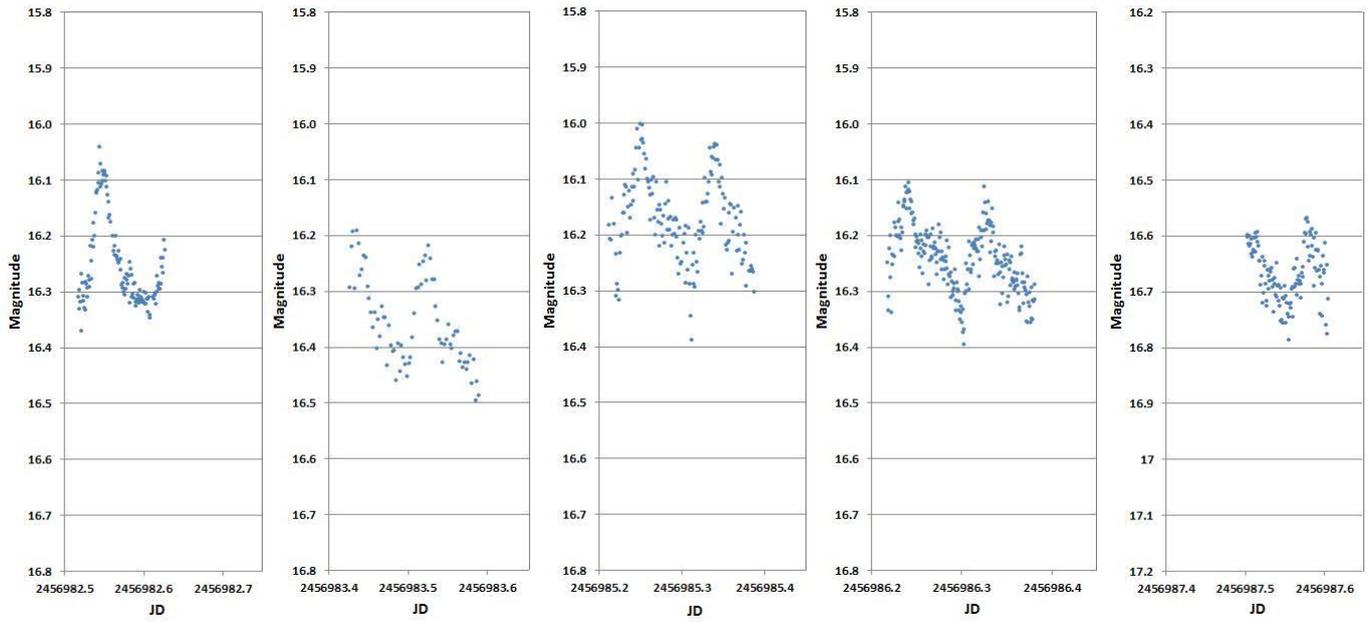

Figure 3: Details of time series photometry during the 2014 November superoutburst showing superhumps

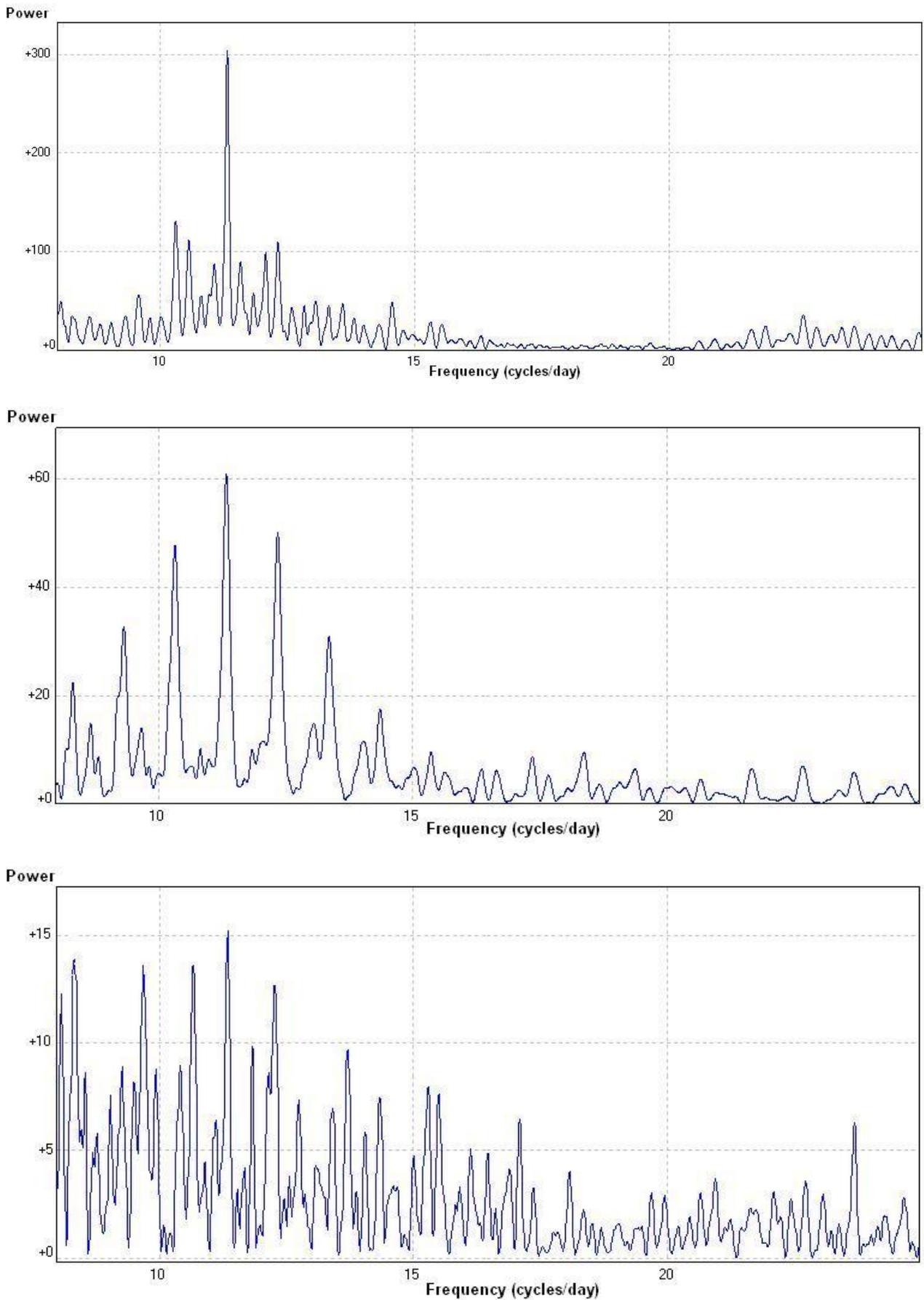

Figure 4: Power spectra of photometric data using the ANOVA method from the 2014 superoutbursts

(a) – top- 2014 November; (b) – middle- 2014 September; (c) – bottom- 2014 July

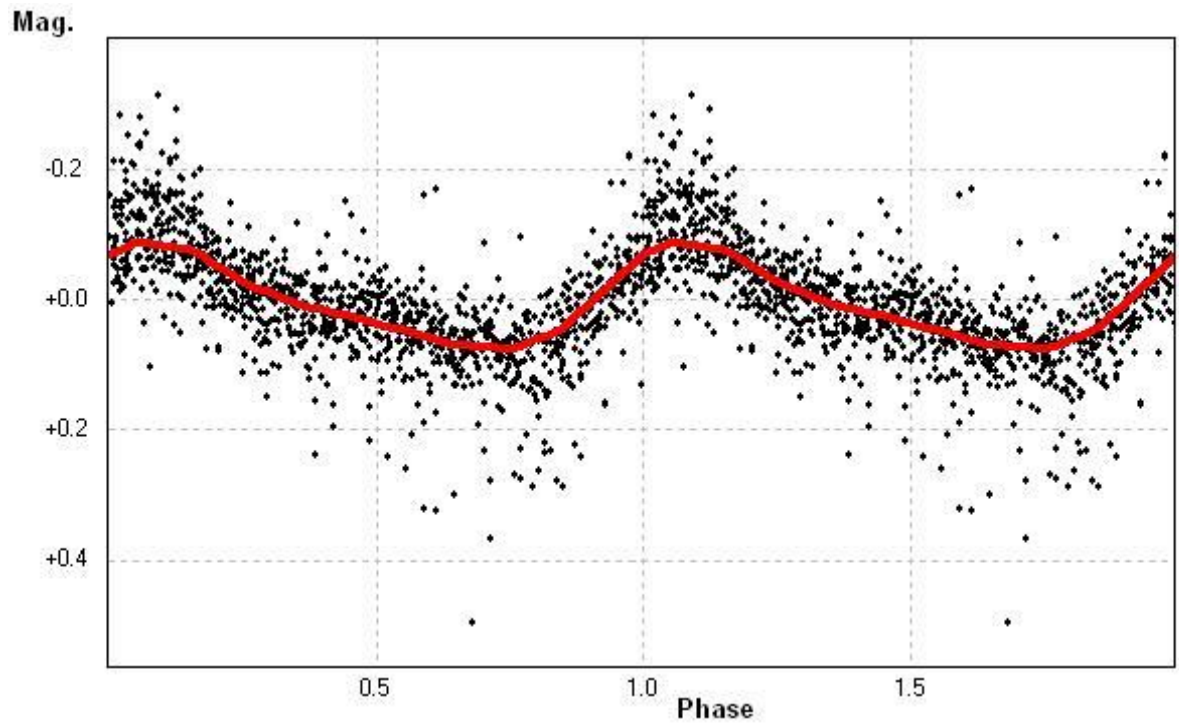

Figure 5: Data from the 2014 November superoutburst folded on the superhump period, $P_{sh}$ = 0.08838(18) d. The red line shows a fourth order polynomial fit to the data using the Polynomial Fit function in *Peranso* (16) to guide the eye

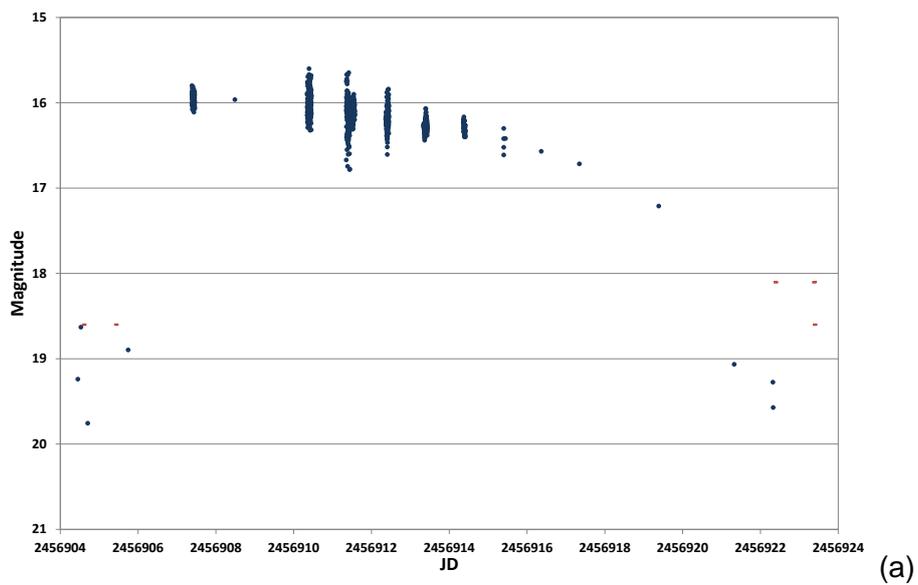

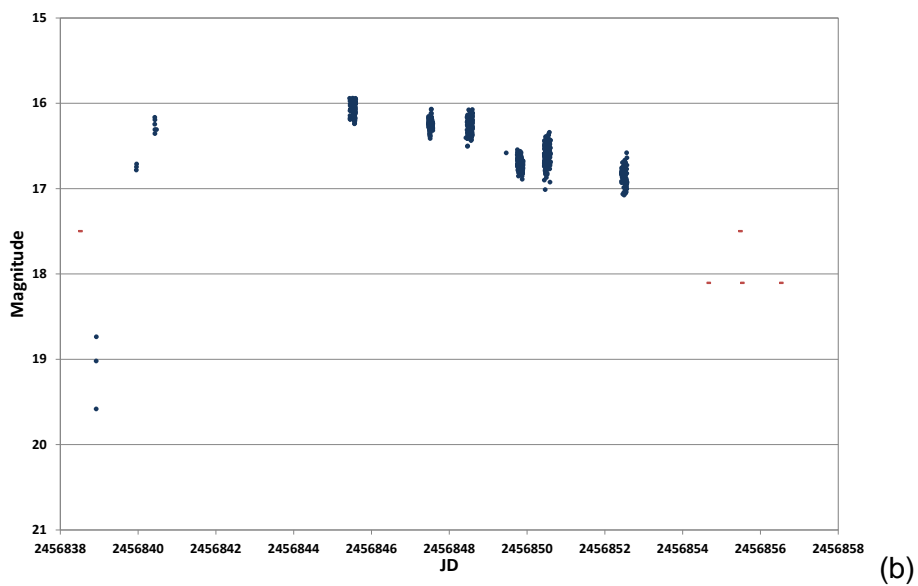

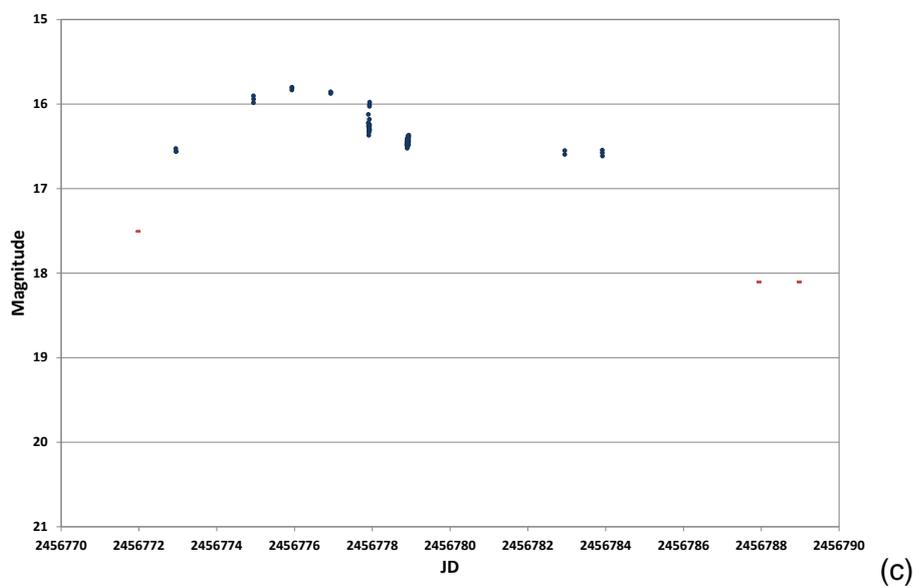

Figure 6: 2014 superoutbursts (a) September, (b) July, (c) April

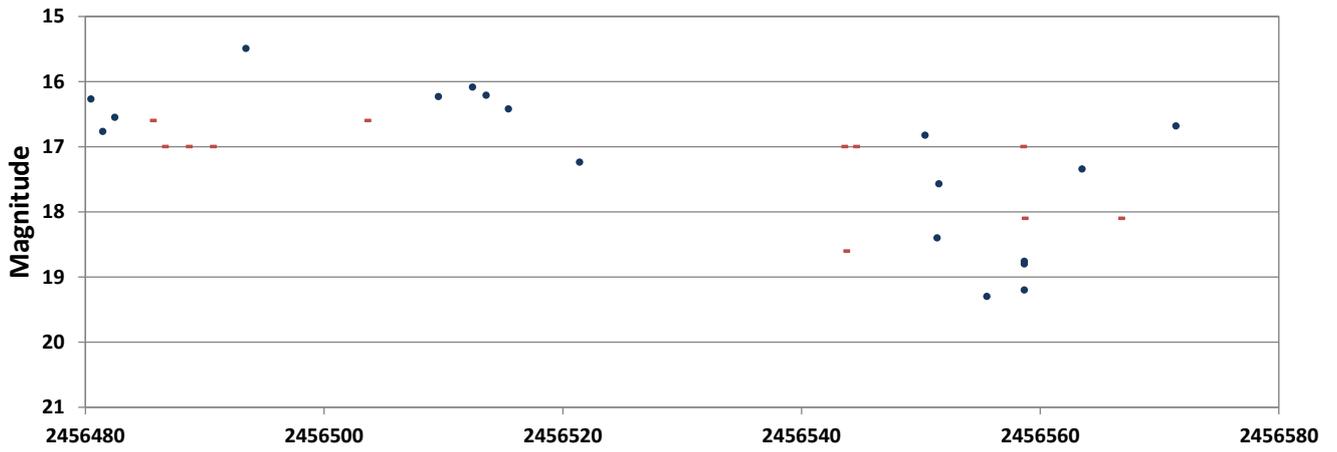

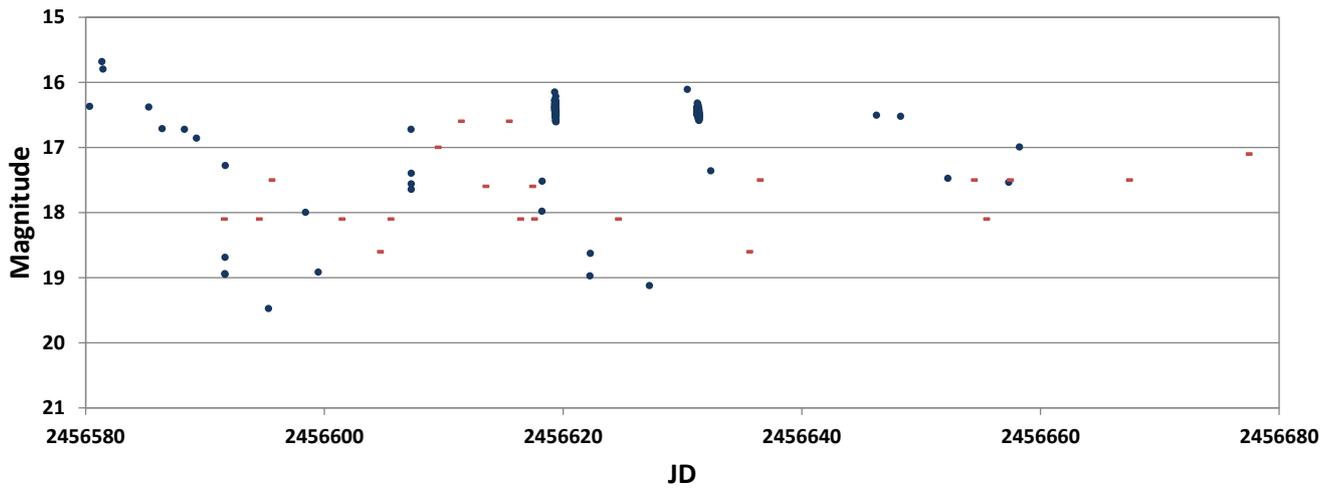

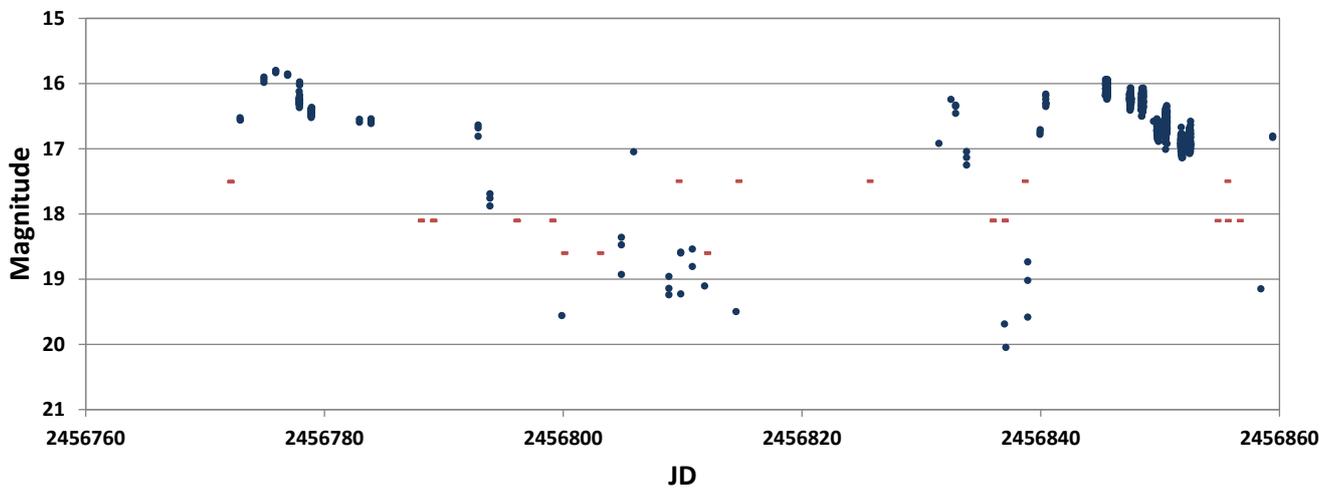

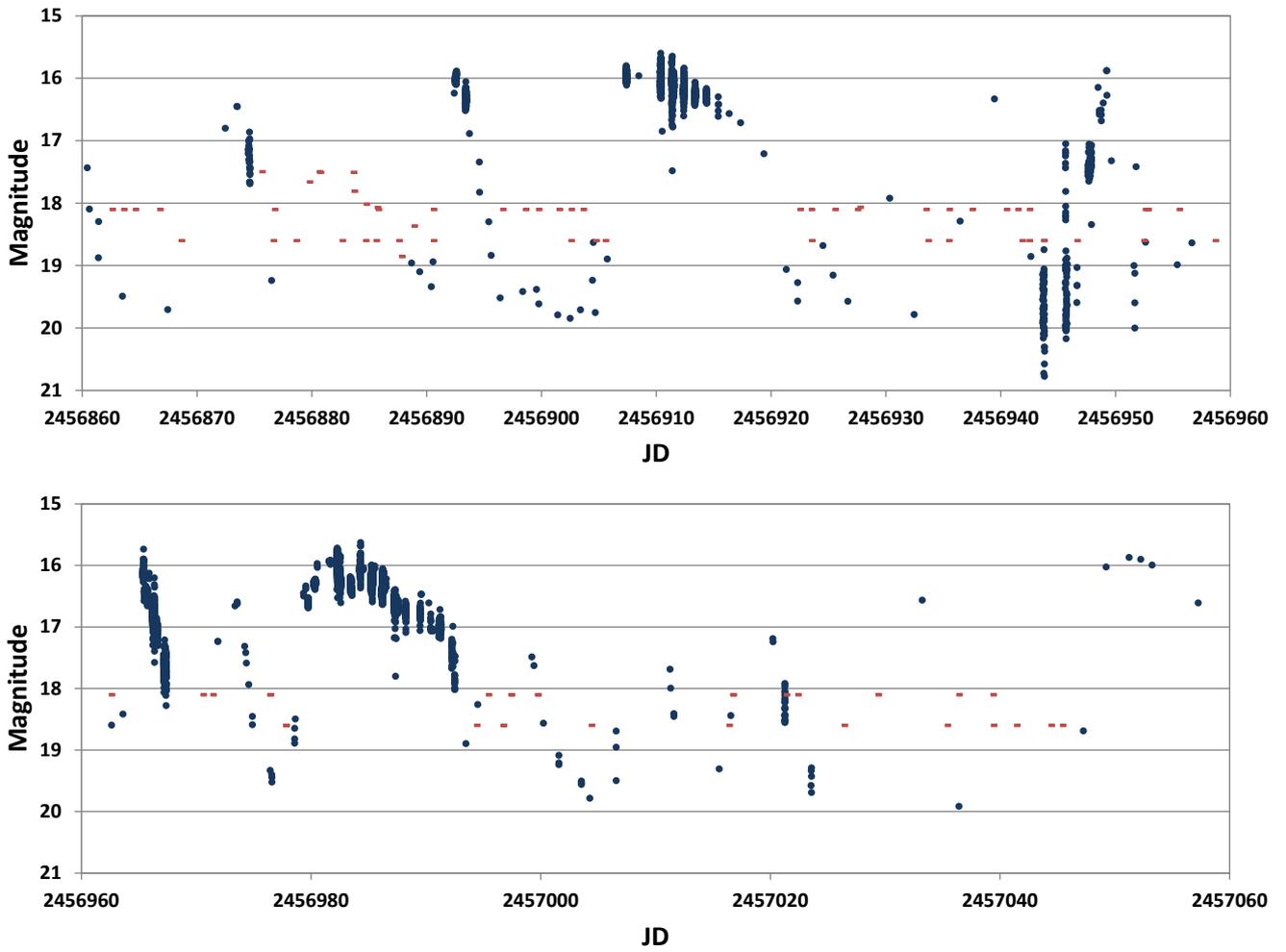

Figure 7: Light curve from 2013 July 6 to 2015 Feb 3

Blue data points are photometric measurements; red dashes are "fainter than" observations

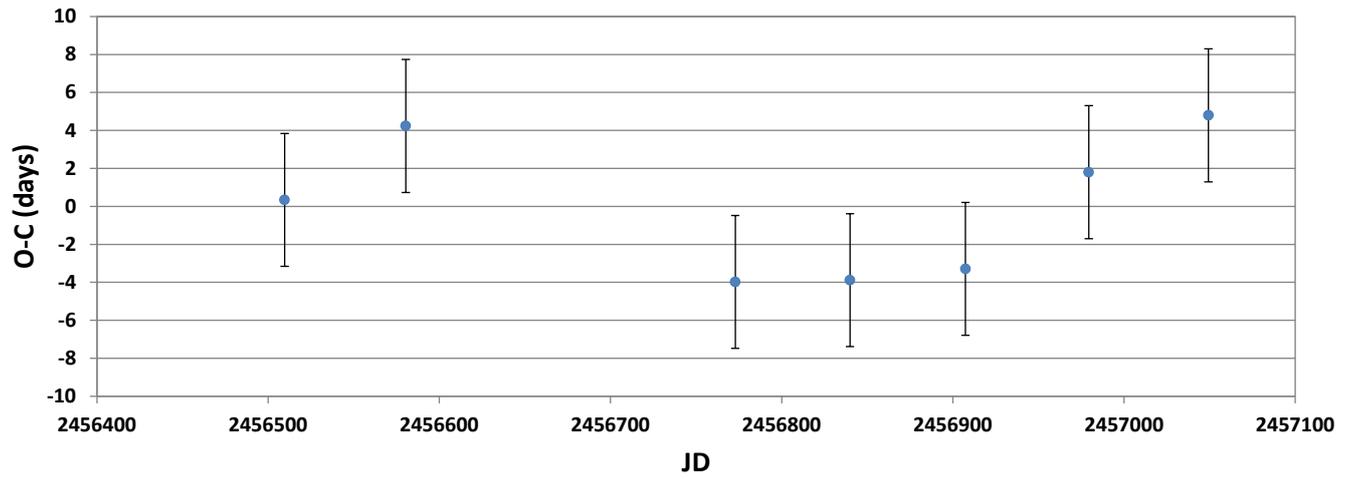

Figure 8: O-C diagram for the superoutbursts using the ephemeris in equation 2 (error bars are RMS of the residuals)